\documentclass[a4paper,fleqn,usenatbib]{mnras}

\usepackage{newtxtext,newtxmath}
\usepackage[T1]{fontenc}
\usepackage{ae,aecompl}

\usepackage{graphicx}	% Including figure files
\usepackage{amsmath}	% Advanced maths commands
\usepackage{amssymb}	% Extra maths symbols
\usepackage[caption=false]{subfig}

\usepackage{epstopdf}
\usepackage{savesym}
\savesymbol{iint}
\usepackage{txfonts}
\restoresymbol{TXF}{iint}

\title[Evolutionary Sequence of Post-Starburst Galaxies]{The Evolutionary Sequence of Post-Starburst Galaxies}

\author[C. L. Wilkinson et al.]{
C. L. Wilkinson,$^{1}$\thanks{E-mail: charlotte.wilkinson@hull.ac.uk (CLW)}
K. A. Pimbblet,$^{1,2}$
J. P. Stott$^{3,4}$
\\
$^{1}$E.A. Milne Centre for Astrophysics, University of Hull, Cottingham Road, Kingston Upon Hull, HU6 7RX, UK\\
$^{2}$School of Physics and Astronomy, Monash University, Clayton, VIC 3800, Australia\\
$^{3}$Department of Physics, Lancaster University, Lancaster LA1 4YB, UK\\
$^{4}$Department of Physics, University of Oxford, Keble Road, Oxford, OX1 3RH, UK
}

\date{Accepted XXX. Received YYY; in original form ZZZ}

\pubyear{2017}

\begin{document}
\label{firstpage}
\pagerange{\pageref{firstpage}--\pageref{lastpage}}
\maketitle

\begin{abstract}
There are multiple ways in which to select post-starburst galaxies in the literature.
In this work, we present a study into how two well-used selection techniques have consequences on observable post-starburst galaxy parameters, such as colour, morphology and environment and how this affects interpretations of their role in the galaxy duty cycle. We identify a master sample of H$\delta$ strong (EW$_{H\delta}$ > 3\AA) post-starburst galaxies from the value-added catalogue in the 7th data release of the Sloan Digital Sky Survey (SDSS DR7) over a redshift range 0.01 < $z$ < 0.1. From this sample we select two E+A subsets, both having a very little [OII] emission (EW$_{[OII]}$ $> -2.5$\AA) but one having an additional cut on EW$_{H\alpha}$ ($> -3$\AA). We examine the differences in observables and AGN fractions to see what effect the H$\alpha$ cut has on the properties of post-starburst galaxies and what these differing samples can tell us about the duty cycle of post-starburst galaxies. We find that H$\delta$ strong galaxies peak in the `blue cloud', E+As in the `green valley' and pure E+As in the `red sequence'. We also find that pure E+As have a more early-type morphology and a higher fraction in denser environments compared with the H$\delta$ strong and E+A galaxies. These results suggest that there is an evolutionary sequence in the post-starburst phase from blue disky galaxies with residual star formation to passive red early-types.
\end{abstract}


%%%%%%%%%%%%%%%%%%%%%%%%%%%%%%%%%%%%%%%%%%%%%%%%%%

%%%%%%%%%%%%%%%%% BODY OF PAPER %%%%%%%%%%%%%%%%%%

\section{Introduction}

Post-starburst galaxies are a vital link between star-forming spirals and quiescent E/S0 galaxies. E+A galaxies, also known as k+a galaxies, are specific type of post-starburst galaxy that were identified by \cite{Dressler1983} as having an elliptical morphology but rich in A-class stars. \cite{Dressler1983} find that 26 E/S0 galaxies display spectra showing no [OII]$_{\lambda3727}$ or H$\alpha$ emission but have deep Balmer absorption lines. The absence of [OII] indicates that there is no ongoing star formation whilst the presence of H$\delta$ absorption is a typical sign of young and recently formed A-class stars. This mix of spectral lines and morphology suggests a recent starburst approximately 1-3 Gyr ago, that has been quenched by an `unknown' process (\citealt{Dressler1982}; \citealt{Couch1987}).

E+A galaxies are thought to be a transitionary phase between star-forming disks and quiescent early-type galaxies. Since their discovery, there have been many studies on E+As, but there is a large variance on how they are defined. Most observational work adopts a lack of [OII] emission and strong absorption in the H$\delta$ Balmer line to classify E+As (\citealt{Poggianti2009}; \citealt{Vergani2010}; \citealt{RodriguezDelPino2014}). Some have combined the Balmer absorption lines such as H$\delta$ with H$\gamma$ and/or H$\beta$ (\citealt{Zabludoff1996}; \citealt{Norton2001}; \citealt{Chang2001v}; \citealt{Blake2004}; \citealt{Tran2004}; \citealt{Yang2004}) in order to maximise their E+A samples.

However, there are a number of studies that use a lack of H$\alpha$ emission (an indicator of star-formation) in their selection, such as \cite{Quintero2004}, \cite{Hogg2006}, \cite{Goto2007f} and \cite{Goto2007}. While this method eliminates dusty star-formers it simultaneously excludes those E+As that contain H$\alpha$ caused by an active galactic nucleus. With different lines being used to select post-starburst galaxies, we ask if this has an impact on their characteristics and what can this tell us about their star formation duty cycle?

In earlier works E+A galaxies have been observed to be blue (\citealt{Dressler1983}; \citealt{Couch1987}). This disagrees with the originally belief that galaxies turn red immediately after a starburst (\citealt{Larson1980}). \cite{Poggianti1999,Poggianti2009} select E+As on the basis of having EW$_{[OII]}$ emission < 5\AA \ and EW$_{H\delta}$ absorption > 3\AA \ and find that there are two populations of E+As; blue and red, although without a discussion of their origin. Further works by \cite{Tran2004} and \cite{Vergani2010}, who select E+A primarily on [OII] emission and Balmer absorption lines, find that their location on a colour magnitude diagram is towards the redder end of the blue cloud and into the green valley. However, \cite{Quintero2004} and \cite{Hogg2006}, who select E+As without H$\alpha$ emission, find that E+As are located towards the bluer end of the red sequence and into the green valley on the colour-magnitude diagram. This suggests that the lack of  H$\alpha$ emission could be the cause of the red E+A galaxies found in \cite{Poggianti1999}.

Work on the morphology of E+A galaxies reveals them to be generally bulge dominated and in some cases having an underlying disk component (\citealt{Tran2004}, \citealt{Quintero2004}). This combination of bulge and disk components replicates the S0 type morphology and reinforces the link between late- and early-type galaxies. This may suggest that the starburst triggering the E+A phase is centrally located, and not in the disc. The latter study opts for different selection methods yet their results on morphology are unanimous: H$\alpha$ has no effect on the morphology of post-starburst galaxies. \cite{Poggianti1999} find that the majority of E+A galaxies contain signs of spiral morphology; this indicates that the time-scale or process responsible for quenching star-formation is different from the process responsible for a morphological transformation. In turn, this could indicate multiple processes involved in triggering the E+A phase. Some have been shown to have a disturbed morphology, which is an indicator of a recent merger (\citealt{Yang2004}). By identifying E+A galaxies we are essentially catching them in the act of transition. Not all starbursts are strong enough to trigger the E+A phase (\citealt{McIntosh2014}), but it is estimated that 30\% of elliptical galaxies have passed through the E+A phase at some point during their evolutionary duty cycle (\citealt{Tran2003a}; \citealt{Goto2005}) making the E+A phase an important and significant part of galaxy evolution.

The properties of E+A galaxies are found to vary with different environments. For example, cluster E+A galaxies contain a more prominent disk component than those outside a cluster (\citealt{Tran2003a}) and are found to be 1.5$\sigma$ bluer than those in the field at $z$ $\sim$ 0.1. This finding goes against the typical colour-density relation that suggests galaxies in clusters should be redder, not bluer (\citealt{Hogg_2003}). This difference in colour is attributed to differing galaxy masses in both field and cluster E+A galaxies. Clusters have been found to contain much smaller E+As than in the field (\citealt{Tran2003a}).

E+A galaxies are found both in a cluster environment as well as a field environment, however, different studies show different environmental preferences. From a sample of 21 nearby E+A galaxies (selected by [OII] and Balmer lines) from the Las Campanas Redshift Survey, \cite{Zabludoff1996} finds 75\% of the E+As are in the field. They find 5 E+A galaxies within their sample display tidal features, a typical sign of galaxy-galaxy interactions and mergers. This suggests that mergers and galaxy-galaxy interactions could be the trigger for the E+A phase. The type of mergers found to dominate in low density environments are gas-rich (\citealt{Bekki2001c}; \citealt{Lin2010}; \citealt{Sanchez-Blazquez2009a}) which are most likely to trigger a starburst strong enough of producing an E+A signature.

When H$\alpha$ is included into the selection criteria different results are found; \cite{Mahajan2013h} find that E+A galaxies prefer a weak-group environment ranging from 4 to 10 galaxies. For local X-ray bright galaxy clusters at 0.02$<z<$0.06, they find 86\% of local clusters contained some form of substructure on a weak group-scale (4-10 members) and 91.4\% of these weak-groups contained E+A galaxies. This result suggests that star-formation is quenched due to a galaxy being pre-processed in a weak-group environment before being accreted into a cluster environment.

Opposite results are found by \cite{Poggianti2009}: at higher redshifts ($0.4<z<0.8$) E+A galaxies are predominantly in clusters where the star-formation has been quenched on a short time scale of $\ll$ 1\,Gyr. \cite{Poggianti2009} find only a few E+A galaxies are located in other environments such as the field and in weak-groups and that the variation in E+A fraction is found to be dependant on environment. \cite{Mahajan2013h}. Regardless of whether H$\alpha$ has been used in the selection criteria of E+As there is a discrepancy in what different works report with regards to environmental preference. \cite{Lemaux2016} suggests that `true' E+As (those with no H$\alpha$) have a higher fraction in a cluster environment compared to those selected by `traditional' methods (using only Balmer absorption lines and the absence of [OII]). Could environment be driving changes in H$\alpha$ emission or are we seeing two different stages of the E+A duty cycle?

A typical sign that star-formation has ceased is a lack of H$\alpha$ emission and as E+As galaxies are no longer producing stars, one would expect to see no significant H$\alpha$ emission. Although not commonly used as a criteria for E+A selection, this work aims to test whether including a cut in H$\alpha$ could explain some of the differences found in other studies and connect up these potentially different phases of post-starburst evolution. Here we present a study of the properties of low redshift post-starburst galaxies using different selection criteria from the Sloan Digital Sky Survey Data Release 7 (SDSS DR7; \citealt{Abazajian2009}) with the broad aim of testing the star formation duty cycle of post-starburst galaxies.

In section 2 we describe the data used in this study and the different selection cuts we impose. In section 3 we analyse and discuss the results. Our conclusions are presented in in section 4. We assume the following cosmological parameters for a flat Universe: $H_0 = \rm 69.3\,kms^{-1} Mpc^{-1}, \Omega_M = 0.238 \ and\ \Omega_{\Lambda} = 0.762$ (\citealt{Spergel2007}).

\section{Sample Selection}
\begin{figure*}
	\centering
	\subfloat[\label{subfig-1:dummy}]{%
		\includegraphics[width=0.49\textwidth]{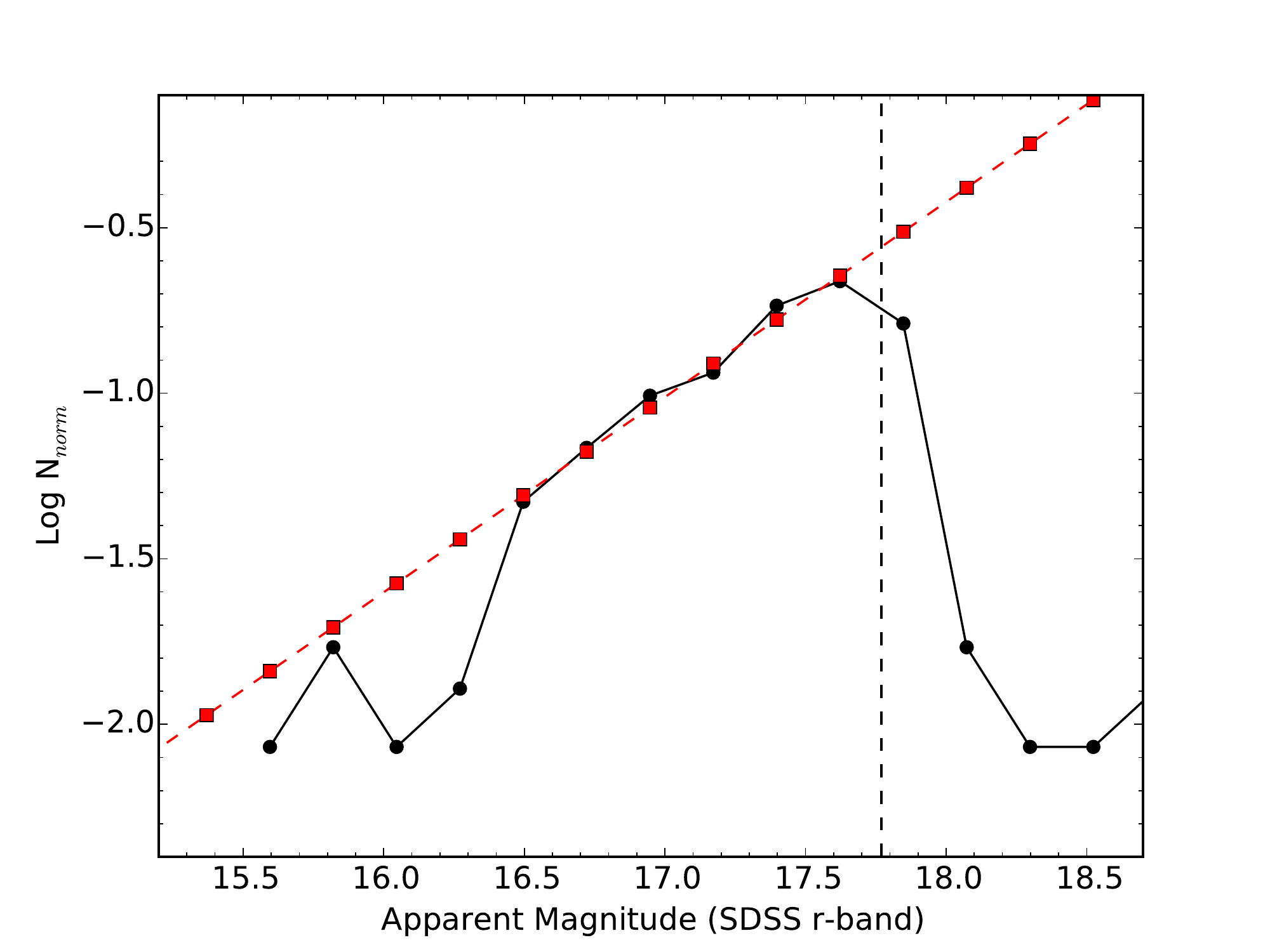}
	}
	\hfill
	\subfloat[\label{subfig-2:dummy}]{%
		\includegraphics[width=0.49\textwidth]{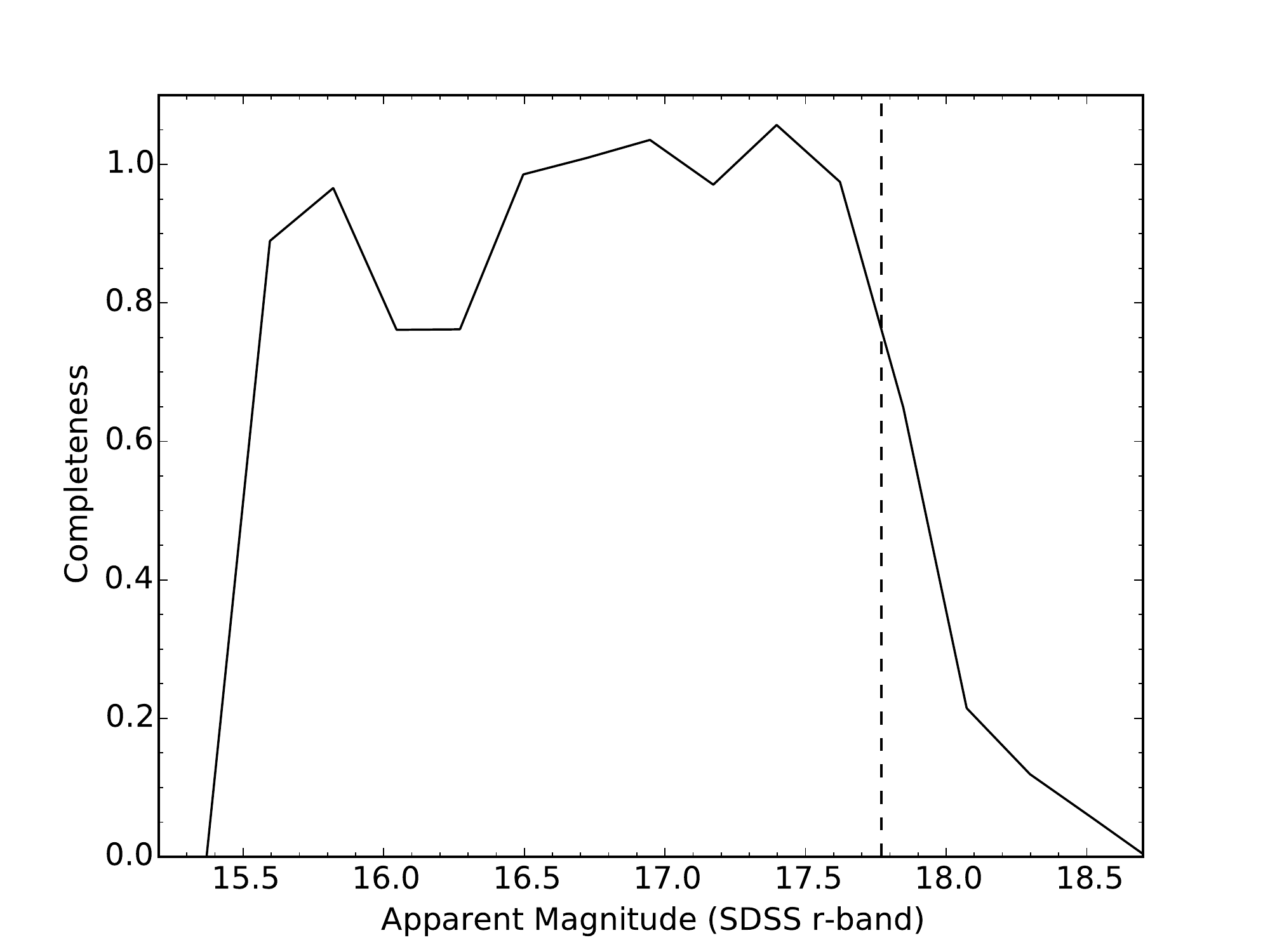}
	}

	\caption{Plot (a) shows the normalised distribution of apparent (model) magnitude in the SDSS r-band of  our H$\delta$ strong galaxy sample. The red dashed line shows the line of best fit which is plotted to the linearly increasing region(16.5$<$r$<$17.5). From this line we obtain plot (b) that shows a completeness diagnostic plot, here we find that we are $\sim$80\% complete to a magnitude limit of r=17.77.}
	\label{magnitude}
\end{figure*}

Using the MPA/JHU value-added catalogue from SDSS DR7 (\citealt{Abazajian2009}), we create a master sample of 234 H$\delta$ strong (EW$_{H\delta}$ > 3\AA; note that positive/negative values for equivalent widths refer to absorption/emission respectively) post-starburst galaxies with 0.01 $< z <$ 0.10. This limit on H$\delta$ is comparable to studies such as \cite{Poggianti1999}, \cite{Bekki2001c}, \cite{Vergani2010} and \cite{RodriguezDelPino2014}. The lower limit on the redshift range ensures we exclude any stars from our sample. Although the SDSS spectroscopy aims to be 100\% complete at r=17.77 (\citealt{Strauss2002}), we find that our sample is $\sim$80\% complete at this level, as shown in Fig.\ref{magnitude}.
%It is here, at r=17.77 that we make our magnitude cut (Fig.\ref{redshift_magnitude}).
Most studies of post-starbursts use a typical S/N of 3-5 in the continuum, as evidenced by \cite{Pracy2005} (S/N$>$5); \cite{Couch1987} (S/N$>$3) and \cite{Brown2009} (S/N$>$4), hence we adopt a minimum S/N of 5. 

Our complete S/N $>$ 5 H$\delta$ strong sample contains 192 galaxies. Using this sample we apply additional cuts to create a further two sub-samples; E+As and `pure' E+As (E+As with low H$\alpha$ emission) in order to test whether different criteria for post-starburst galaxies can explain the differences found in literature and to gain a better insight into the post-starburst duty cycle.

%\begin{figure}
%	\centering
%	\includegraphics[width=0.5\textwidth]{redshift_magnitude.pdf}
%	\caption{This scatter plot shows the distribution of apparent (model) magnitudes with redshift. The dashed line represents the cut we make at r=17.77, this ensures we select a complete sample H$\delta$ galaxies.}
%	\label{redshift_magnitude}
%\end{figure}

Our first sub-sample is based on the \cite{Poggianti1999} definition of E+A galaxies. With the primary selection of EW$_{H\delta}$ > 3\AA, we apply an additional cut of $>-2.5$\AA\ in the equivalent width of [OII]$_{\lambda3727}$ (\cite{Zabludoff1996}; \citealt{Yang2004}). This sample of E+As contains 67 galaxies.

To account for this, our second sub-sample ensures there is no ongoing (potentially dusty) star-formation in the E+A sample by applying a further cut in H$\alpha$ emission (EW$_{H\alpha}$ > -3\AA) (\citealt{Goto2007}). Doing this we obtain 25 pure E+A galaxies. We list the selection criteria for our three post-starburst sub-samples below:

\noindent H$\delta$ Strong:
\begin{enumerate}
	\item EW$_{\delta}$ > 3\AA
\end{enumerate}
E+A:
\begin{enumerate}
	\item EW$_{\delta}$ > 3\AA
	\item EW$_{[OII]}$ > -2.5\AA
\end{enumerate}
Pure E+A:
\begin{enumerate}
	\item EW$_{H\delta}$ > 3\AA
	\item EW$_{[OII]}$ > -2.5\AA
	\item EW$_{H\alpha}$ > -3\AA
\end{enumerate}

\begin{figure}
	\centering
	\includegraphics[width=0.5\textwidth]{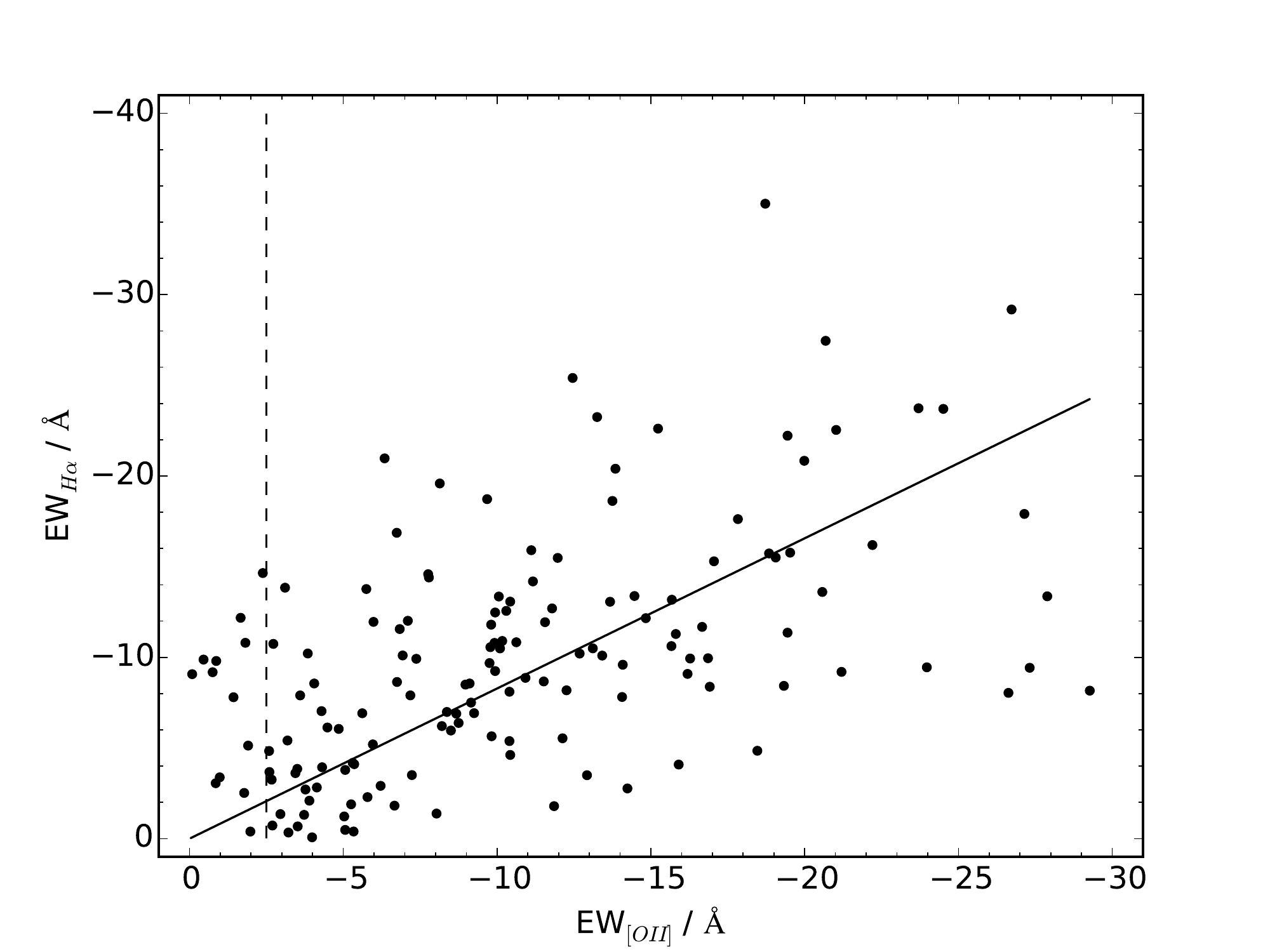}
	\caption{This plot shows the equivalent widths of [OII] and H$\alpha$. Note that positive/negative values for equivalent widths refer to absorption/emission respectively. The [OII] limit has been highlighted by the dashed line. The trend line (solid) has been fitted using a robust linear regression model (Huber 1981). This plot shows a significant amount of scatter in which R$^2$=0.30.}
	\label{EQWs}
\end{figure}

\begin{figure*}
\centering
\subfloat{
	\includegraphics[width=1\textwidth]{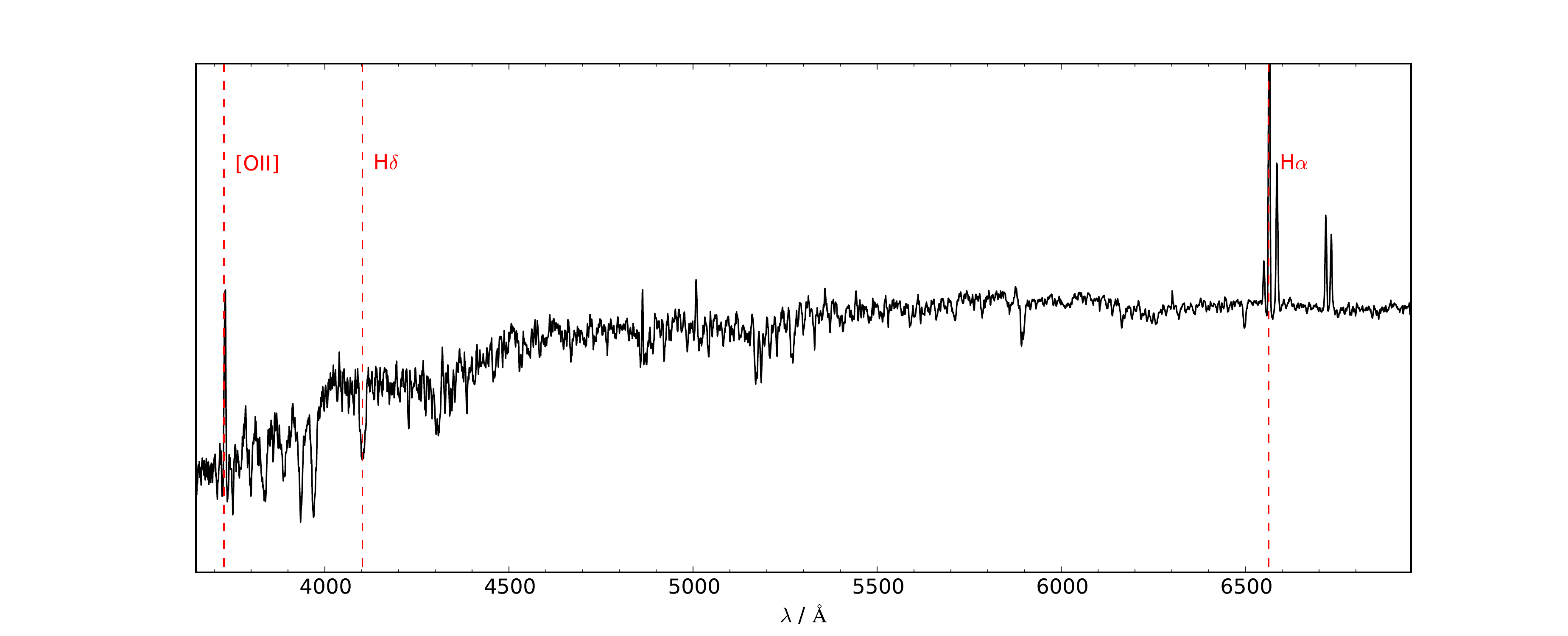}
}\\
\subfloat{
	\includegraphics[width=1\textwidth]{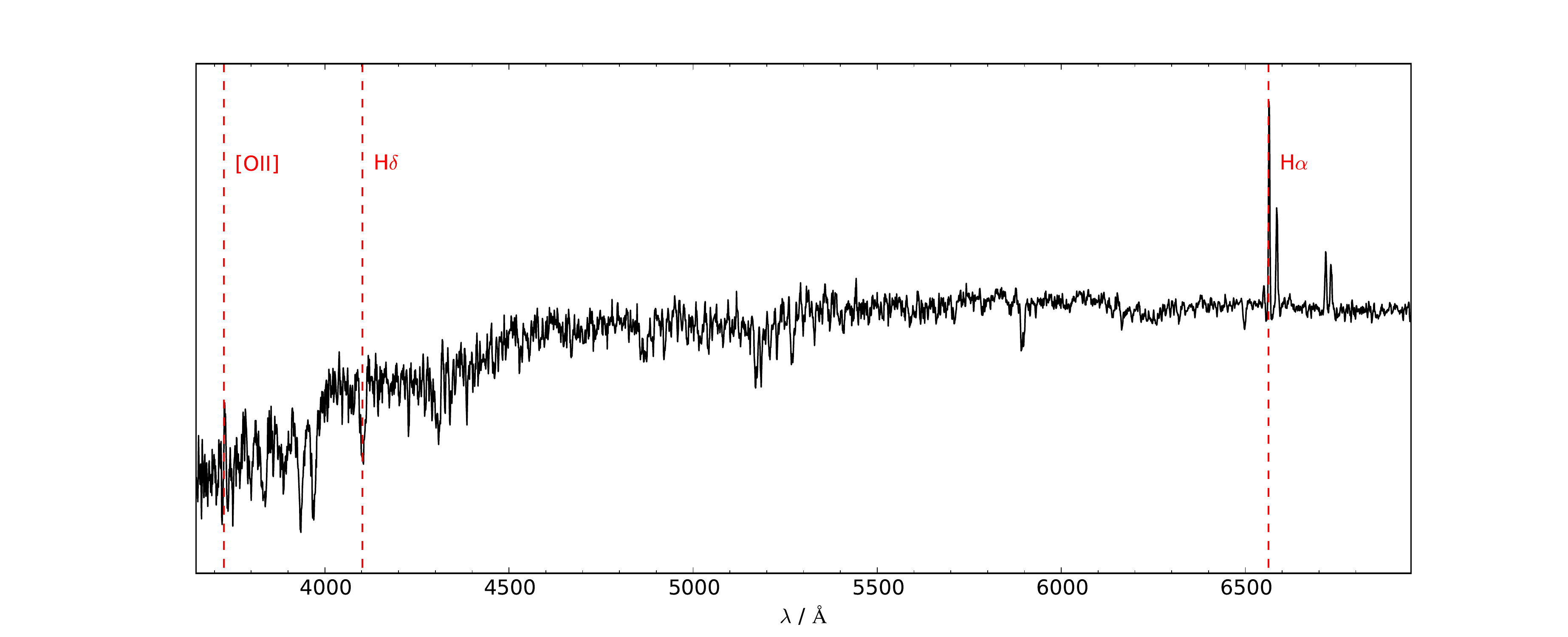}
}\\
\subfloat{
	\includegraphics[width=1\textwidth]{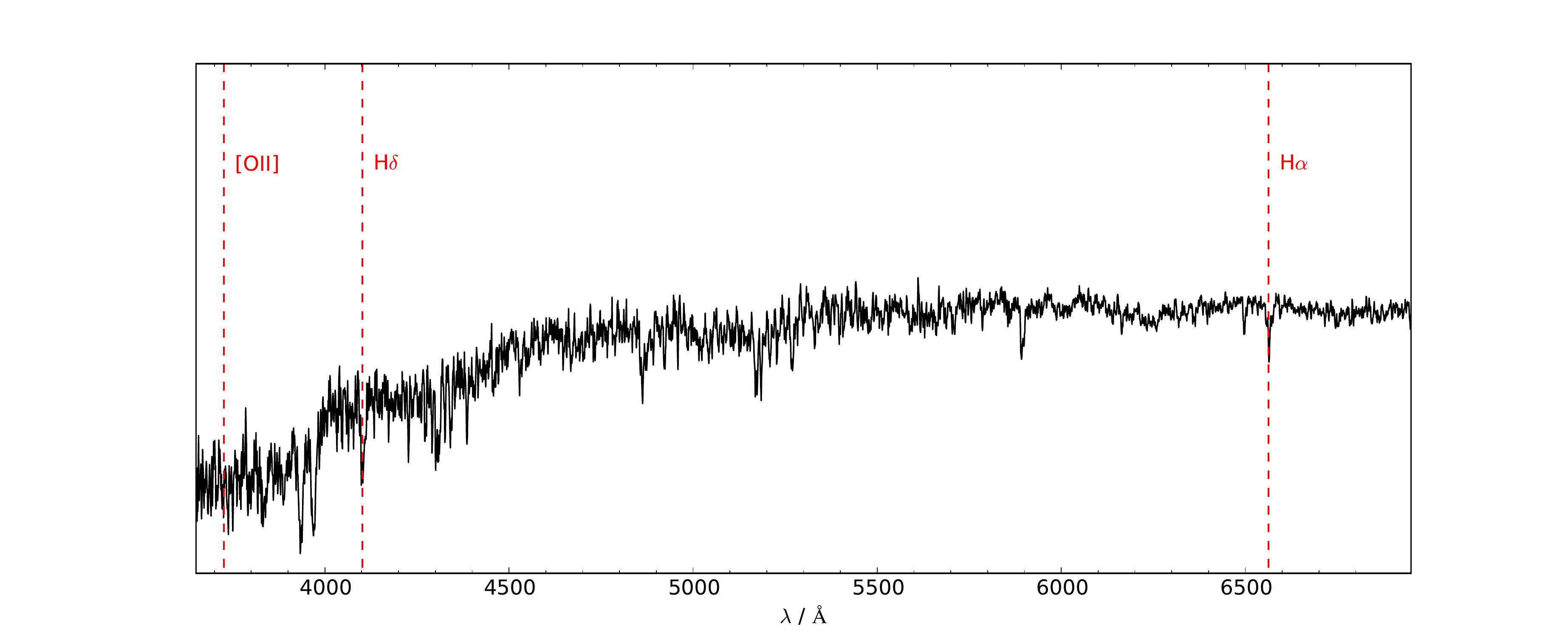}
}
\caption{These are stacked median spectra for each of the populations of post-starburst galaxies in this study; H$\delta$ strong (top), E+A (middle) and pure E+A (bottom). We have included the locations of the three spectral lines used for selection in this study (dashed lines). The y-axis is arbitrary flux units.}
\label{example_spec}
\end{figure*}

We test whether or not we would expect to see H$\alpha$ given the limit on [OII] emission stated above. We use a simple approach by plotting EW$_{[OII]}$ against EW$_{H\alpha}$  as shown in Fig. \ref{EQWs}. In this plot we witness a significant amount of scatter (regression analysis gives R$^2$ to be 0.3), however, there is an underlying trend present and we are able to fit a regression line using a robust linear model (\citealt{Huber1981}). From this we see at the EW$_{[OII]}$ limit we would expect EW$_{H\alpha}$ to be -2.1\AA\ in emission. On average the [OII] cut should remove any significant star formation indicated by H$\alpha$. However, the large scatter means that there is a population of galaxies with up to EW=-15\AA\ of H$\alpha$ emission at the [OII] limit (See fig. \ref{EQWs}). We also calculate the star formation rate derived from the H$\alpha$ flux, SFR$_{H\alpha}$ from \cite{Kennicutt1998} and apply an extinction correction (10$^{A/-2.5}$) from \cite{Treyer_2007}. Assuming $Av=0.5$ we find the average SFR$_{H\alpha}$ for H$\delta$ strong galaxies that didn't make the E+A cut to be 0.07M$_{\odot}$/yr$^{-1}$. For the E+A that are not classed as pure we find SFR$_{H\alpha}$ to be 0.06M$_{\odot}$/yr$^{-1}$ and pure E+A have an average SFR$_{H\alpha}$ of 0.01M$_{\odot}$/yr$^{-1}$. Stacked spectra for our three sub-samples can be found in Fig. \ref{example_spec}. We note here that emission-filling could potentially affect our H$\delta$ measurements. We believe this would be a small effect (\citealt{Goto2003b, Blake2004}). However, the usual method of measuring this in conjunction with, e.g., H$\alpha$ to determine the strength of its effect is unable to be achieved due to our sample selection approach.

%\begin{table}
%	\centering
%\begin{tabular}{c|c|c|c}
%	\hline
%	&A=0.0&A=0.5&A=1.0\\
%	\hline
%	Predicted EW$_{H\alpha}$&2.960\AA&5.404\AA&9.850\AA\\
%	Predicted EW$_{[OII]}$&11.878\AA&17.333\AA&25.284\AA\\
%	\hline
%\end{tabular}
%\caption{This table shows the predicted equivalent widths for H$\alpha$ and [OII] emission lines for A=0.0, 0.5 and 1.0 at the [OII] and H$\alpha$ limits respectively.}
%\label{line_predictions}
%\end{table}

%\begin{figure}
%	\centering
%	\includegraphics[width=0.5\textwidth]{Robust_OII.pdf}
%	\caption{This plot shows the emission equivalent widths of OII and H$\alpha$, in which the trend line is fitted using robust linear regression (Huber 1981). There is a large amount of scatter in this plot in which R$^2$=0.24}
%	\label{OII}
%\end{figure}
%\begin{figure}
%	\centering
%	\includegraphics[width=0.5\textwidth]{Robust_Ha.pdf}
%	\caption{This plot shows the emission equivalent widths of OII and H$\alpha$, in which the trend line is fitted using robust linear regression (Huber 1981). We quantify the scatter to be R$^2$=0.52.}
%	\label{Ha}
%\end{figure}

\section{Colour}
In this section we investigate whether the differences in selection criteria, based on spectral lines, has an impact on the colours of the post-starburst population. This will allow us to determine why \cite{Poggianti1999,Poggianti2009} finds two populations of E+A galaxies that peak in the red sequence and blue cloud.

\begin{figure*}
	\centering
	\includegraphics[width=1.1\textwidth]{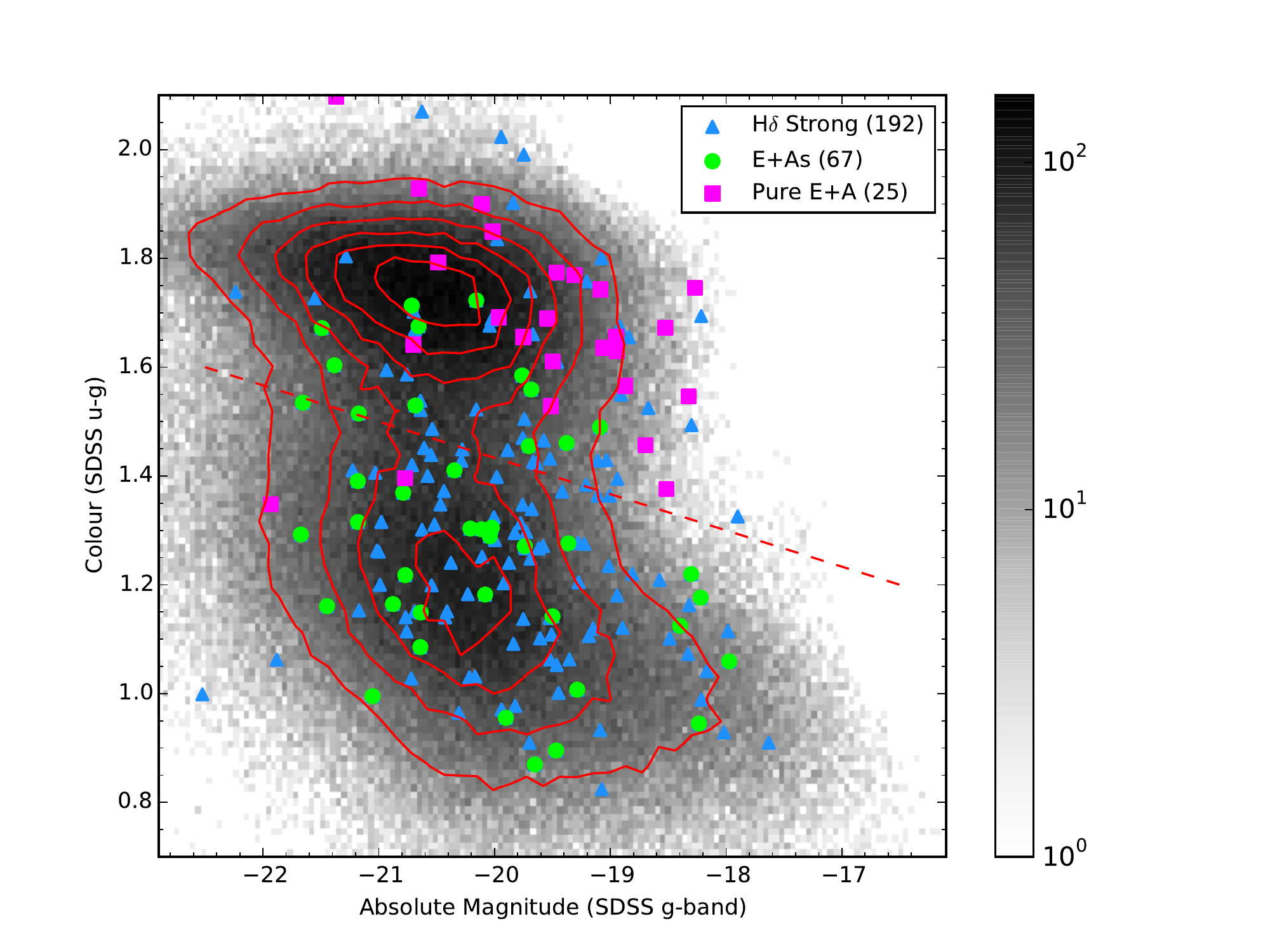}
	\caption{This plot shows a colour magnitude diagram with contour lines (red) that show the regions of the red sequence and blue cloud separated by a dashed line to guide the eye. These regions are based on the SDSS DR7 data shown in grey. The vast majority of pure E+As present predominantly lies in the red sequence whilst the remaining sample reside mainly in the blue cloud. We have included the number of galaxies in each sub-sample in brackets within the legend.}
	\label{colour_mag}
\end{figure*}

To study the differences in colour between our samples we first plot a colour magnitude diagram, shown in Fig. \ref{colour_mag}. Using the galaxies from the SDSS DR7 catalogue, we are able to locate the red sequence and blue cloud regions and over plot our samples to see which regions they reside (Fig.\ref{colour_mag}).

\begin{figure*}
	\centering
	\includegraphics[width=1.1\textwidth]{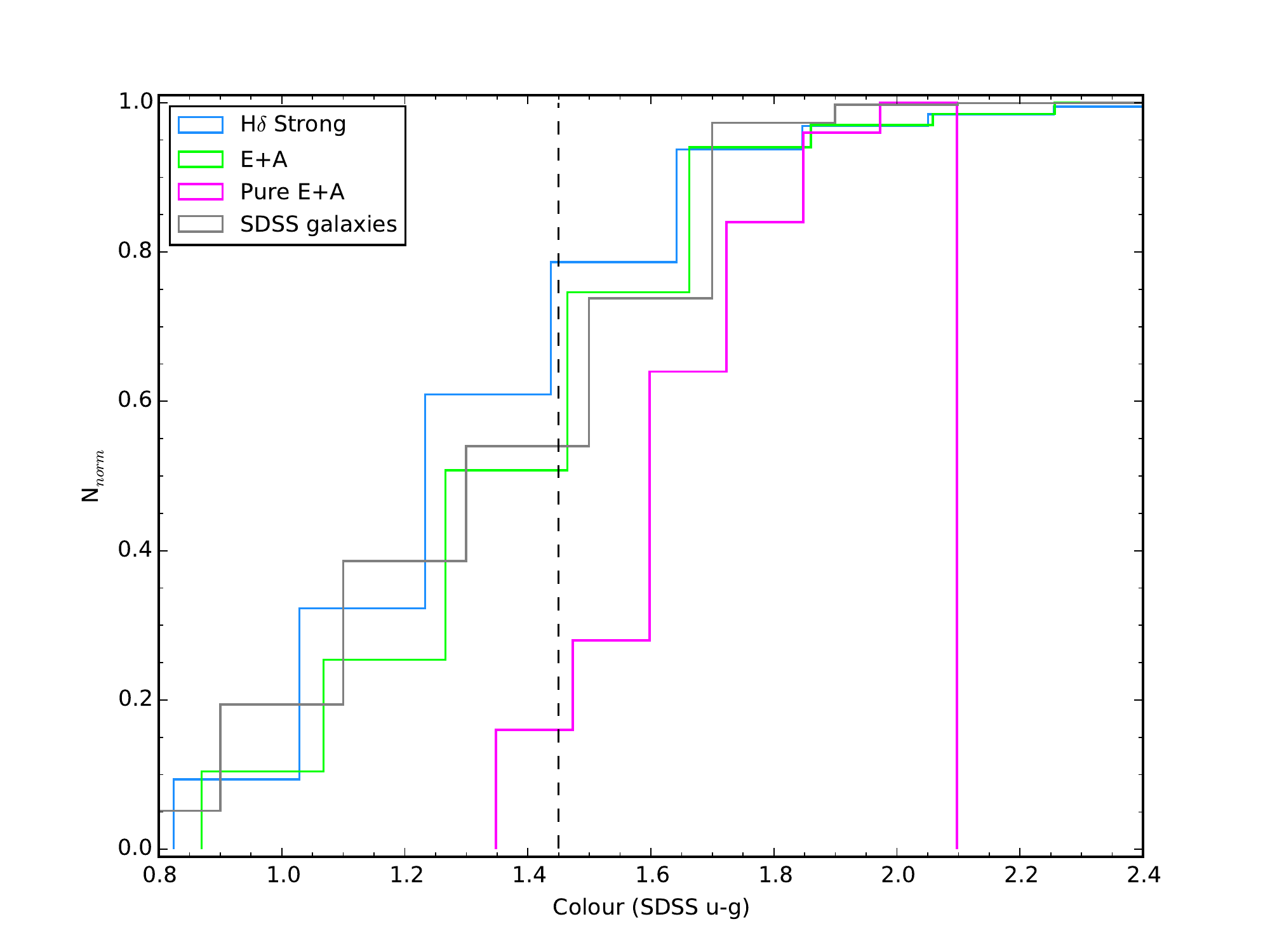}
	\caption{This histogram shows the cumulative distributions of colour between the three samples in this study and SDSS galaxies. From this we see the H$\delta$ strong galaxies peak in the blue cloud and the E+As peak in the green valley. The pure E+As are well established in the red sequence.}
	\label{colour_hist}
\end{figure*}

We include a colour histogram (Fig. \ref{colour_hist}) that shows the bi-modality of the colour distribution. Using Fig. \ref{colour_mag} and Fig. \ref{colour_hist} we find that the H$\delta$ strong sample peaks in the blue cloud with a median colour of 1.36$\pm0.04$. The E+A sample peaks in the green valley with a median colour of 1.46$\pm0.04$. We find that the pure E+As predominantly reside in the red sequence with a median colour of 1.66$\pm0.04$.

\cite{Hogg2006} select E+A galaxies using a cut in H$\alpha$ and find that E+As are located in the red sequence towards the green valley. While E+As have a significant connection to the green valley (\citealt{Vergani2010}), we find that by selecting E+As using H$\alpha$ emission we are predominantly selecting those towards the red sequence as star formation has ceased and the galaxies are truly passive.

To test whether the E+A and pure E+A samples are distinguishable we apply a two sample Kolmogorov-Smirnov statistical (abbreviated to KS onwards) test. When comparing the E+A sample with the pure E+As the KS test returns a p-value of 0.007 which demonstrated that E+As and pure E+As are unlikely to be drawn from the same parent distribution. In this study we consider any p-value less than 0.01 to be statistically different, whilst p-value over 0.1 we consider to be the same. P-values between these limits we consider to have a weak difference.

To test the distinguishability between the E+A sample and the H$\delta$ strong sample, we apply the same test and output a p-value of 0.110. These distributions are likely to be from the same parent distribution. When we take H$\delta$ strong galaxies that are not defined as E+As and compare to the E+A sample we get a p-value of 0.005. This means there is a statistical difference between non-E+A H$\delta$ strong galaxies and E+A galaxies. Fig. \ref{colour_hist} clearly shows that the pure E+A sample and the H$\delta$ strong sample are statistically different. To quantify the distinguishability of the H$\delta$ strong sample and pure E+As we include a p-value which is found to be 10$^{-6}$.

\section{Morphology and Morphological Parameters}
To explore the differences in findings on morphology we test the morphological fractions of each sample using the Galaxy Zoo (GZ) catalogue (\citealt{Lintott2008,Lintott2011}). The GZ project provides an extensive catalogue of galaxy morphologies and is based on the data from SDSS. GZ is a citizen science project in which volunteers view images of galaxies and cast their vote as to whether the galaxy is elliptical, spiral, merger or unknown. We assign a classification of elliptical, spiral or merger to our galaxies if that classification takes 50\% or more of the votes. The minimum number of votes per galaxy in our sample is 16 and the mean number or counts our sample has is 39.5. We note here that S0 galaxies are not easily distinguishable from the elliptical and spiral morphologies in the GZ catalogue. Whilst the majority of S0 galaxies are within the elliptical classification there will certainly be S0 contamination in the spiral classifications (\citealt{Bamford2009}).

We find that from our H$\delta$ strong sample we are able to match 185 (96.4\%) galaxies to the GZ catalogue. This in turn means we matched 62 E+As (92.5\%) and 21 pure E+As (84.0\%) to the catalogue. The morphology fractions for each of the samples are shown in Table \ref{morphologies}. We find that there is no statistically significant merger fraction in either of the three samples, this is due to the very low numbers that were identified. We find that $\sim$55\% of the H$\delta$ strong sample have a spiral morphology and $\sim$30\% have an elliptical morphology. There is a clear majority here for a preference of disc morphologies. This result somewhat changes in the E+A sample in which there is a $\approx$50/50 (42\%/44\%) split between the elliptical and spiral morphologies for those matched to GZ. However, in the pure E+A sample we see $\sim$80\% of the galaxies are classed as ellipticals and no significant spiral fraction. This is in agreement with the findings that pure E+As are redder than the H$\delta$ strong and E+A galaxies and occupy the red sequence with the majority of early-type. Interestingly we note that mergers make up less than 5\% of any of the post-starburst categories and therefore do not seem to be important for this phase of galaxy evolution.

\begin{figure}
	\centering
	\includegraphics[width=0.5\textwidth]{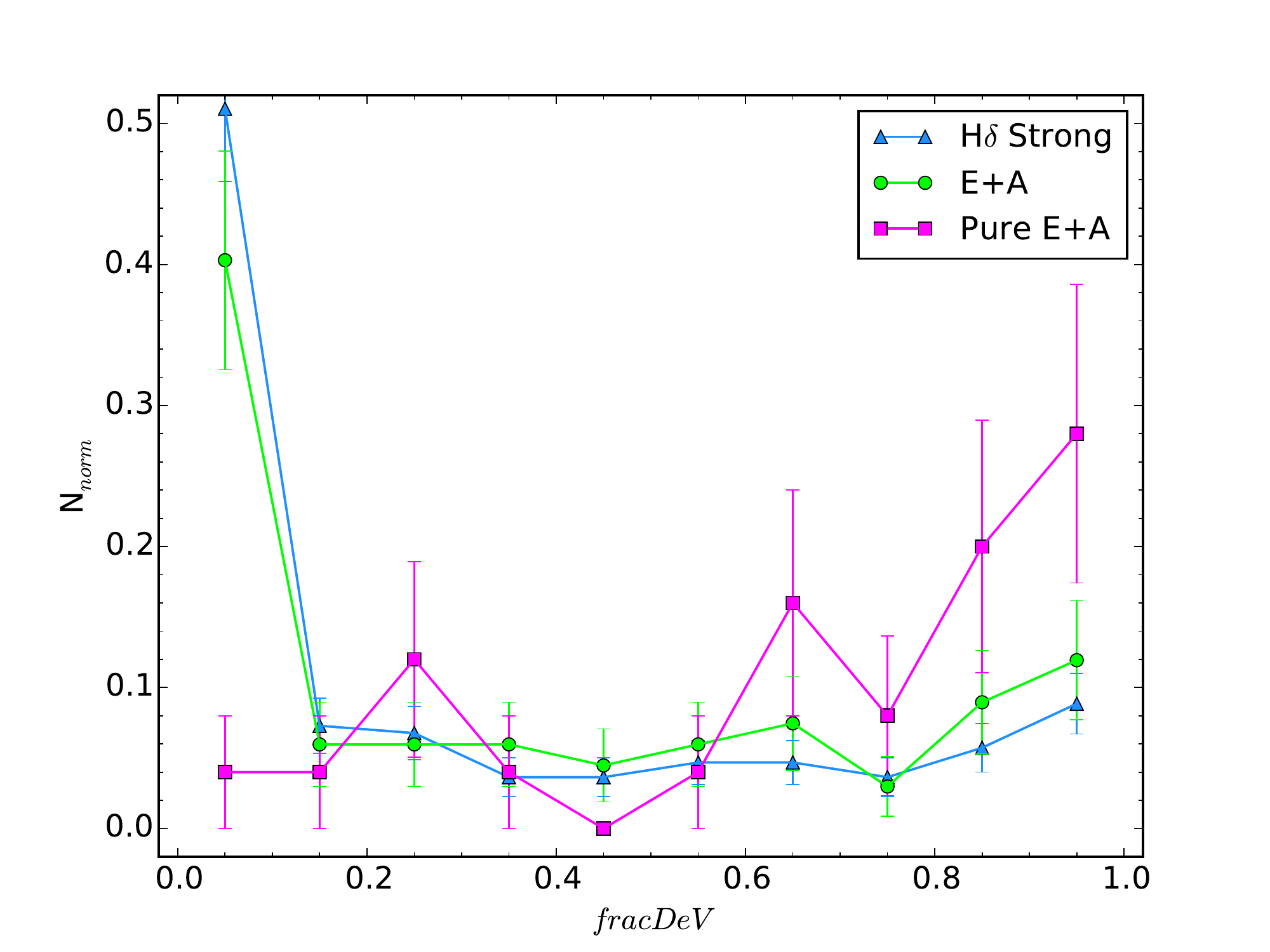}
	\caption{This plot shows the normalised distributions of the $fracDeV$ parameter for the three samples in this study. We see that the H$\delta$ strong galaxies and E+A galaxies tend to a disk-like light profile whereas the pure E+As tend to an elliptical profile.}
	\label{fracdev}
\end{figure}

As well as using the GZ catalogue to test morphology fractions we also look at the $fracDeV$ parameter, which corresponds to the S\'{e}rsic index (a generalization of the de Vaucouleur profile, \citealt{Vaucouleurs1948}; \citealt{Sersic1968}) and links together galaxy shape with surface brightness profiles (Fig. \ref{fracdev}). $fracDeV$ is the fraction of light fit by the de Vaucouleur profile versus an exponential profile. When $fracDeV$=1, this represents a pure de Vaucouleur profile and is appropriate for an elliptical morphology. When $fracDev$=0, this represents a pure exponential profile and is appropriate for a disk morphology as seen in spiral galaxies. The $fracDev$ parameter is listed in the SDSS database for all $u,g,r,i,z$ filters (\citealt{Abazajian2009}).

Fig. \ref{fracdev} shows the normalised distribution of the $fracDeV$ parameter and broadly agrees with the morphological classifications in Table \ref{morphologies}. For the pure E+A the $fracDeV$ distribution tends to one meaning an elliptical morphology. For the H$\delta$ strong sample and the E+A sample we visually see little difference in distribution and that both distributions are dominated by a disk-like profile. We calculate their median $fracDeV$ values to be 0.086$\pm0.025$ for H$\delta$ strong galaxies, 0.238$\pm0.046$ for E+As and 0.779$\pm0.063$ for pure E+As.To test this statistically we apply a KS test to all samples so determine their distinguishability. We find that when comparing the H$\delta$ strong sample to the E+A sample we obtain a p-value of 0.338; we conclude that these two samples are not statistically different. To determine whether the E+A sample and the pure E+As emission are distinguishable we apply the same test and obtain a p-value of 0.003. We can therefore claim that there is a significant difference between E+As and pure E+As. To complete these tests, a p-value of $\sim$10$^{-6}$ was achieved when comparing the H$\delta$ strong sample to the pure E+As.

We examine the radii of our samples to further quantify our morphology results and to compare to the trends found in \cite{Shen2003}. Here we use the effective radii based on the deVaucouleur profile in the SDSS r-band. We note here we have included a first order seeing correction to the radii, as shown in Equation \ref{seeing} assuming the average SDSS seeing of 2". 

\begin{equation}\label{seeing}
r_{corrected} = (r_{observed}^2 - seeing^2)^{1/2}
\end{equation}

\begin{figure}
	\centering
	\includegraphics[width=0.5\textwidth]{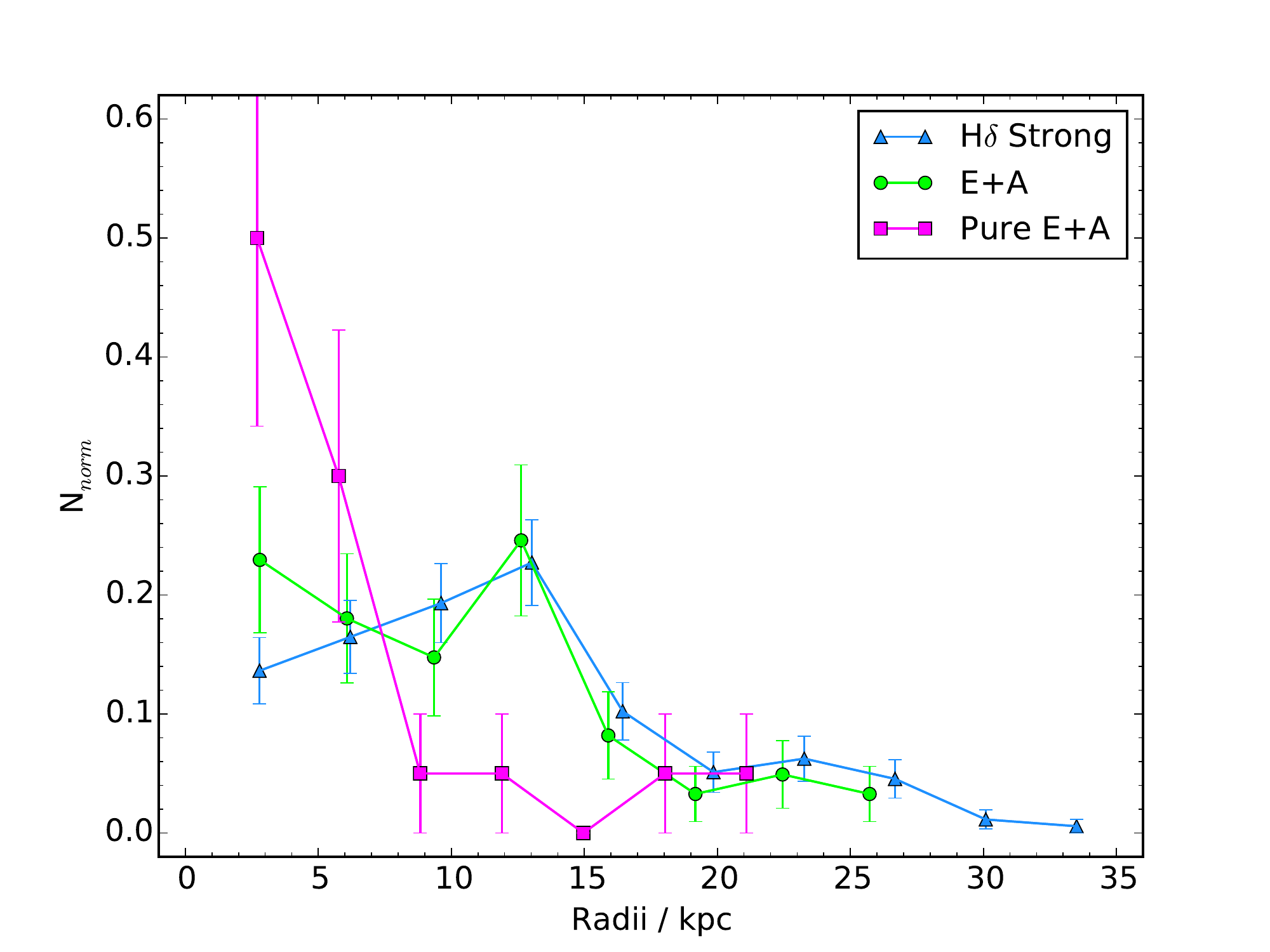}
	\caption{This plot shows this normalised distributions of radii for the three samples. The H$\delta$ strong sample has a median radii of 11.4kpc$\pm$0.5, whereas the E+A population has a median radii of 9.8kpc$\pm$0.8. The median radii of the pure E+As is 4.1kpc$\pm$1.2.}
	\label{radii}
\end{figure}

\begin{figure}
	\centering
	\includegraphics[width=0.5\textwidth]{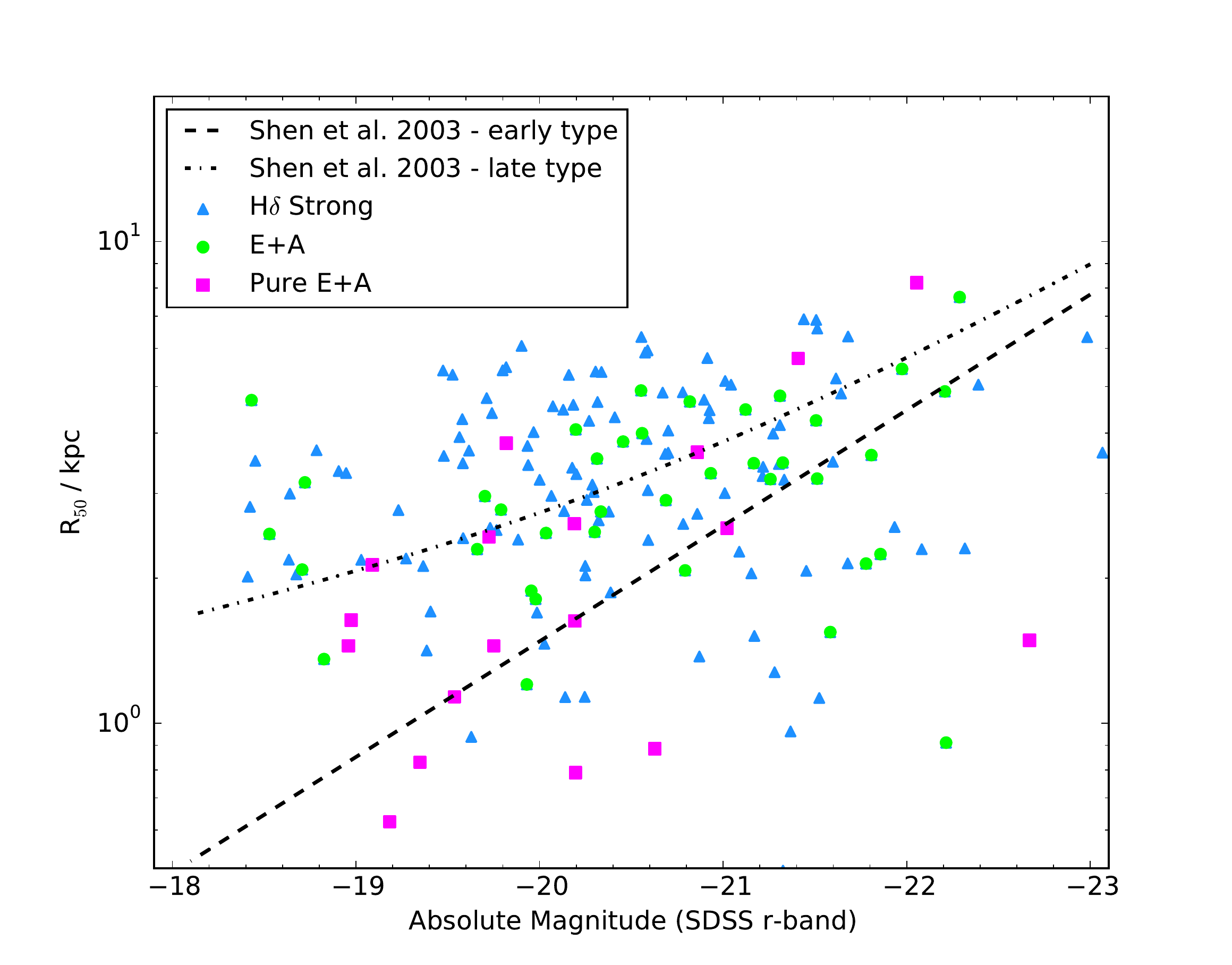}
	\caption{This plot shows the relation between absolute magnitude and the half light radius, $R_{50}$. For comparison we include the relationships found in Shen et al. (2013). We find that our H$\delta$ strong and E+A galaxies follow the late-type trend whereas the pure E+As follow the early-type trend.}
	\label{shen_mag}
\end{figure}

Fig. \ref{radii} shows the normalised radius distributions of the three samples. The majority of the pure E+A galaxies are mainly under 10kpc in radius and are therefore relatively small galaxies. We find the following median radii for the H$\delta$ strong, E+A and pure E+A galaxies respectively: 11.4kpc$\pm$0.5, 9.8kpc$\pm$0.8 and 4.1kpc$\pm$1.2. When a KS test is applied to the H$\delta$ strong sample and the E+A sample we obtain a p-value of 0.176. When comparing the E+As and H$\delta$ strong sample to the pure E+As we obtain the p-values 0.006 and $\sim$10$^{-6}$ respectively. This results shows that our third sample are statistically smaller in radii than the other two samples. This is also shown in Fig. \ref{shen_mag} which shows the relation between radius and absolute magnitude with the corresponding trends from  \cite{Shen2003}.  We look at the median separation between R$_{50}$ and the trends from \cite{Shen2003} to quantify which line fits which population. From the H$\delta$ strong sample we find that the median separation from the early-type trend is 1.21kpc and 0.03kpc from the late-type trend. Looking at the E+A sample we find that the median separation is 0.48kpc from both the early-type trend and late-type trend, this demonstrate the finds from Table \ref{morphologies}. Finally, we find that the median separation of the pure E+A sample to the early-type trend is 0.39kpc and 0.85kpc to the late-type trend. This shows the H$\delta$ strong galaxies follow the trend for late-type galaxies, the E+As an even mix of early- and late-type and the pure E+As resemble an early-type relation. This again confirms our findings from the colour and morphology analysis.

\begin{table*}
	\centering
	\begin{tabular}{c c c c c}
		\hline
		Sample & Elliptical & Spiral & Merger & Unclassified\\
		\hline
		H$\delta$ Strong & 0.297$\pm$0.040 & 0.541 $\pm$0.054 & 0.005$\pm$0.005 & 0.157$\pm$0.029\\
		E+A & 0.419$\pm$0.082 & 0.435$\pm$0.084 & 0.016$\pm$0.016&0.130$\pm$0.044\\
		Pure E+A & 0.810$\pm$0.206 & 0.048$\pm$0.048 &0.048$\pm$0.048&0.094$\pm$0.061\\
		\hline
	\end{tabular}
	\caption{This table shows the morphology fractions for each of the samples along with the Poisson error. These morphologies were assigned using the galaxy zoo catalogue in which a particular classification took 50\% or more of the votes. Here we see the majority of H$\delta$ strong galaxies take up a spiral morphology, whilst E+As are equally found as ellipticals and spirals. We see that pure E+As are predominantly elliptical.}
	\label{morphologies}
\end{table*}

%\begin{figure}
%	\centering
%	\includegraphics[width=0.5\textwidth]{mass_err.pdf}
%	\caption{This plot shows this normalised distributions of mass for the three samples. All three samples follow very similar mass distributions. We see a dip in the distribution at Log(M/M$_{\odot}$) = 9.7, this could be due to a selection effect however the reason is unclear.}
%	\label{mass}
%\end{figure}

%\begin{figure}
%	\centering
%	\includegraphics[width=0.5\textwidth]{log_Shen_mass_seeing.pdf}
%	\caption{This plot shows the relation between mass and the half light radius. For comparison we include Shen lines that show the spirals and ellipticals in our sample are relatively large.}
%	\label{shen_mass}
%\end{figure}

\section{Environment}
Whilst there is little evidence in the literature that selection criteria affect the environmental results, we aim to test whether there are any environmental differences between our sub-samples.

In order to explore the environments of the galaxies in our samples we use the Yang group and cluster catalogue (\citealt{Yang2007}). We directly match 176 galaxies in our sample (91.2\%); this corresponds to 64 E+As (94.0\%) and 22 pure E+As (88.0\%). The Yang catalogue is based on the 4th data release of SDSS and selects galaxies with a redshift completeness $C > 0.7$, so therefore may not match totally to the 7th data release.

First, we examine whether our galaxies are in groups/clusters, if so, we look at the membership of that group or cluster to determine the size. Our results are displayed in Table \ref{yang}. We split our environment classifications into field galaxies, pairs, weak groups (3$<$N$<$10), rich groups (10$<$N$<$50) and clusters (N$>$50).

Table \ref{yang} shows that $\sim$70\% and $\sim$60\% of the H$\delta$ strong galaxies and E+A galaxies respectively are field galaxies with a strong preference for this type of environment; this supports the findings of \cite{Zabludoff1996} who find that 75\% of E+A galaxies to be in a field environment. We find that only$\sim$40\% of the pure E+As are in the field environment whilst there is a significant amount of pure E+A galaxies in the group environments. When combining these environments together into all field (field and pair, where a pair is two galaxies that are gravitationally bound) and all cluster (weak groups, rich groups and clusters) environments we see the following splits in preference (field/cluster); 80/20, 70/30 and 50/50 for H$\delta$ strong, E+A and pure E+As respectively. This suggests that pure E+As are more likely to be in denser environments than the other sub-samples of post-starburst galaxies.

%For those galaxies in our sample that are not  in a `field' or pair environment, we examine whether post-starburst galaxies are more or less likely to be satellite galaxies or are currently in-falling into a group/cluster by examining their position on a phase space diagram (Fig. \ref{phase_space}) and comparing to a caustic. The caustic used here is based on the methodology used by \cite{Carlberg1997}, which selects galaxies greater than 3$\sigma$ away from the mean as being infalling. We see from Fig. \ref{phase_space} that all of our galaxies are within the caustic and are therefore established in their group/cluster and are not currently infalling. As the time scale of the post-starburst phase is significantly shorter that the time scale of migration onto a group/cluster, we can state that the post-starburst galaxies found in groups/clusters were triggered in their current environment.

\begin{table*}
	\centering
	\begin{tabular}{c c c c c c c c}
		\hline
		Sample & Field & Pair & Weak Group & Rich Group & Cluster & All Field & All Cluster\\
		\hline
		H$\delta$ Strong & 0.69$\pm$0.06 & 0.08$\pm$0.02 & 0.11$\pm$0.03 & 0.09$\pm$0.02 & 0.02$\pm$0.01 & 0.77$\pm$0.07&0.23$\pm$0.04\\
		E+A & 0.59$\pm$0.10 & 0.10$\pm$0.04& 0.14$\pm$0.05 & 0.13$\pm$0.05 & 0.05$\pm$0.03& 0.68$\pm$0.11&0.32$\pm$0.07\\
		Pure E+A & 0.41$\pm$0.14 & 0.09$\pm$0.06 & 0.27$\pm$0.11 & 0.14$\pm$0.08 & 0.09$\pm$0.06 & 0.50$\pm$0.15& 0.50$\pm$0.15\\
		\hline
	\end{tabular}
        \caption{This table shows the fractions of galaxies from each sample that reside in different environments. In this work we class field galaxies to be those that are not associated with any group or cluster in the Yang catalogue. We class weak groups as containing between 3 and 10 galaxies, rich groups 10-50 galaxies and clusters  50+ galaxies. The all field category includes galaxies which are in field and pair environments. The all cluster environment includes galaxies which are in groups with 3 or more members.}
	\label{yang}
\end{table*}

%\begin{table*}
%	\centering
%	\begin{tabular}{c c c c c}
%		\hline
%		Sample & N$_{Cluster}$ & N$_{All Group}$&Inside Caustic & Outside Caustic\\
%		\hline
%		H$\delta$ Strong & 4 & 36 & 27 (67.5\%) & 13 (40.0\%)\\
%		E+A & 3 & 17 & 13 (65.0\%) & 7 (35.0\%)\\
%		`Pure' E+A& 2 & 9 & 7 (63.6\%) & 3 (27.3\%)\\
%		\hline
%	\end{tabular}
%\caption{This table shows the number of galaxies that are located within and outside of the caustic in Fig. \ref{phase_space}. The galaxies in this table are all those that are located within a group (weak and rich) and cluster environments.}
%\label{Caustic_Table}
%\end{table*}

%\begin{figure}
%	\centering
%	\includegraphics[width=0.5\textwidth]{phase_space.pdf}
%	\caption{This plot shows a phase space diagram, along with the 3$\sigma$ caustics (dashed line) from Carlberg et al. 1997. Group and cluster members are shown as grey crosses. All of the post-starburst galaxies in this plot are within the caustic, this means they are not in the process of infalling onto a group/cluster.}
%	\label{phase_space}
%\end{figure}

\section{AGN Connection}
Hydrodynamical simulations show that major mergers of spirals contribute heavily to the populations of E+A galaxies (\citealt{Snyder2011a}). When these gas-rich spirals merge, high levels of star-formation occur and to quench this star-formation a feedback mechanism is required, potentially AGN driven feedback.

During a merger gas is funnelled into the nucleus of a merging galaxy by tidal torques and and can trigger a nuclear starburst (\citealt{Mihos1996}; \citealt{Hopkins2008}; \citealt{Hopkins2008a}) and therefore a potential progenitor of the E+A phase. With an excess of gas in the nucleus of the galaxy, this provides fuel for the supermassive black hole in the centre and can trigger an active galactic nuclei (AGN). AGN can produce jets, winds or radiation that heat or remove cold gas from galaxies, removing the fuel for further star formation.

\cite{DePropris2014} show in their study of 10 K+A galaxies that at any age there is no presence of AGN. They suggest, however, that a quasar phase immediately follows the quenching mechanism and this leads onto the K+A phase.

\begin{figure}
	\centering
	\includegraphics[width=0.5\textwidth]{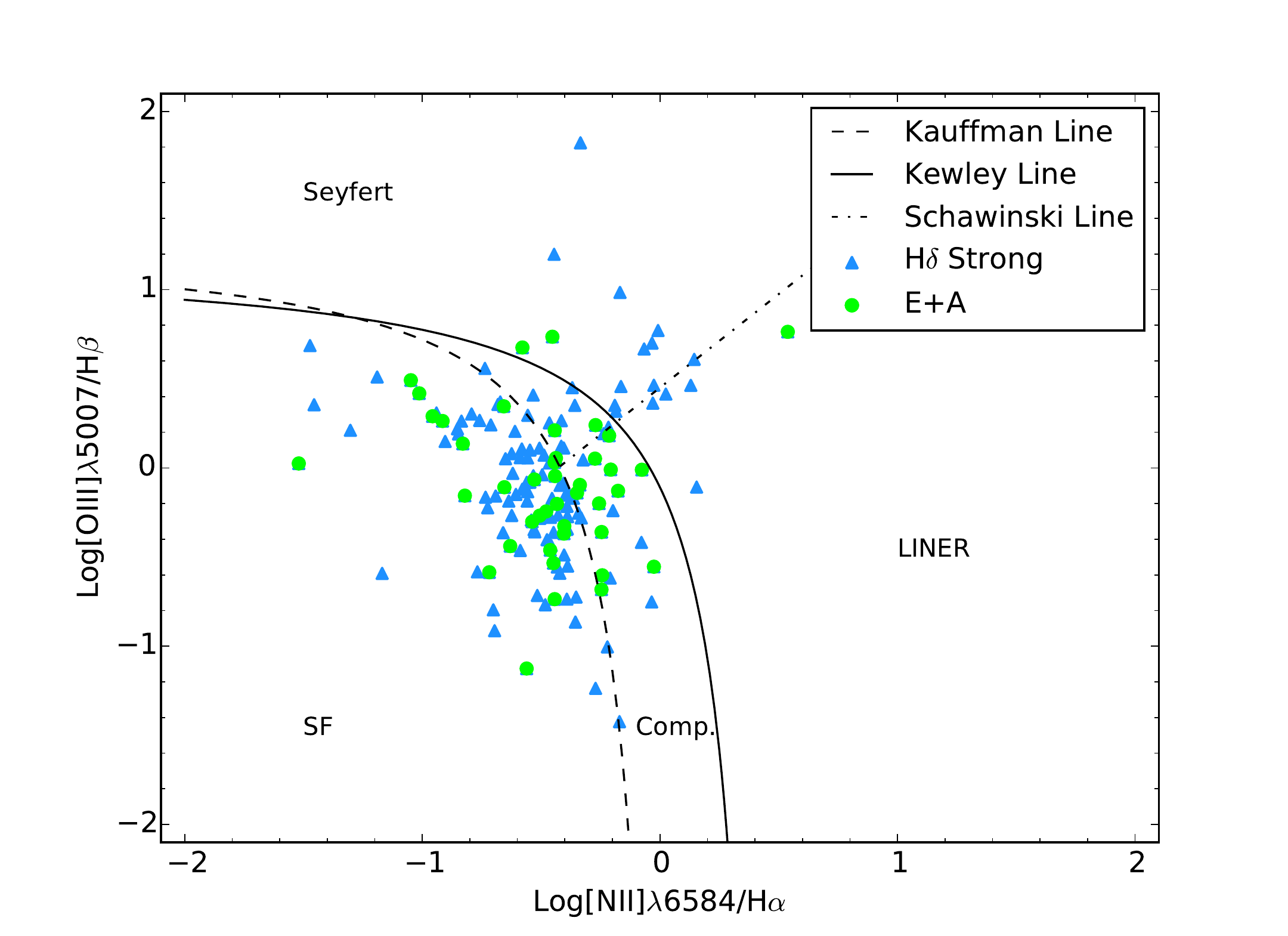}
	\caption{This BPT shows the classification of our samples using the Kauffman (Kauffmann et al. 2003), Kewley (Kewley et al. 2001) and Schawinski lines (Schawinski et al. 2014). We find the majority of the H$\delta$ strong galaxies are located within the star forming region. We find that the E+A galaxies tend towards the star-forming and composite regions.}
	\label{BPT}
\end{figure}

In this section we investigate the E+A/AGN link further by examining the AGN fractions within our samples and comparing to the fractions found in SDSS. We start by plotting a BPT diagram (\citealt{Baldwin1981}; \citealt{Veilleux1987}) shown in Fig. \ref{BPT} which includes the Kauffman (\citealt{Kauffmann2003}), Kewley (\citealt{Kewley2001a}) and Schawinski lines (\citealt{Schawinski2014}). Using these lines we locate the seyfert, LINER, composite and star formation regions. Starting with a control sample derived from SDSS we find that seyfert galaxies make up 10.8\% of the general galaxy population, LINERs 11.5\%, composites 20.7\% and star formers 57.0\%.

We plot the H$\delta$ strong and E+A samples in Fig. \ref{BPT} on the BPT diagram and find that the majority these two samples are within the star forming region (65.4\%$\pm$6.4 and 57.1\%$\pm$11.7 respectively). We find that composites makes up 23.0\%$\pm$3.8 of the H$\delta$ strong sample and 35.7\%$\pm$9.2 of the E+A sample. Seyferts make up 8.2\%$\pm$2.3 of the H$\delta$ strong sample and 4.8\%$\pm$3.4 of the E+As whilst LINERs make up 3.1\%$\pm$1.4 of the H$\delta$ strong sample and 2.4\%$\pm$2.4 of the E+As. These results show similar fraction in both samples of star formers compared to the general galaxy population, however we witness an enhancement in E+A composite galaxies and decline in E+A AGN.

\begin{figure}
	\centering
	\includegraphics[width=0.5\textwidth]{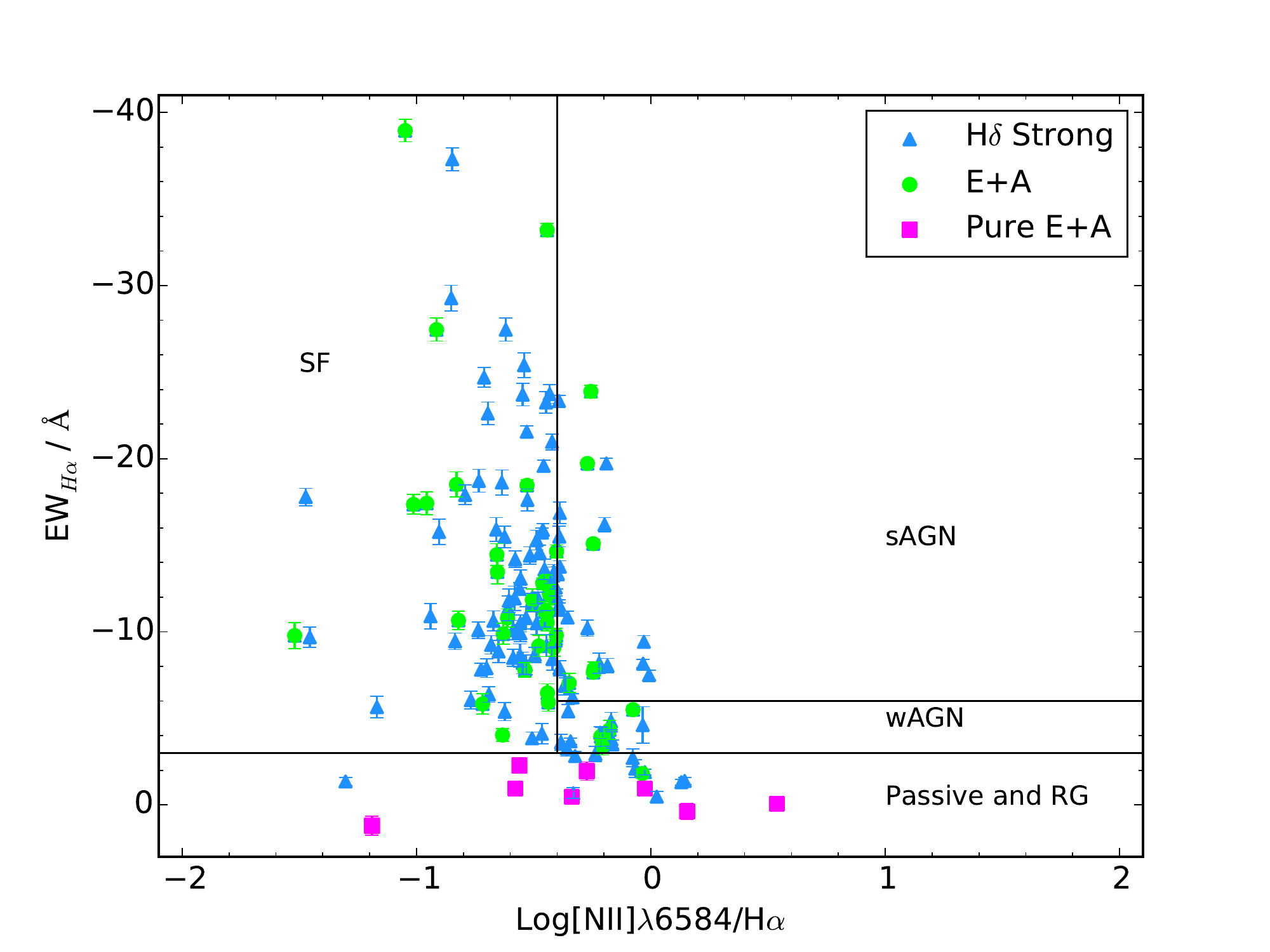}
	\caption{This WHAN diagram shows the majority of the H$\delta$ strong and E+A galaxies are in the star-forming region (59.7\%$\pm$6.1) with a small fraction residing in the AGN sections of the WHAN diagram (25.2\%$\pm$4.0). Error bars are included on all points for the H$\alpha$ line. Note that positive/negative values for equivalent widths refer to absorption/emission respectively.}
	\label{WHAN}
\end{figure}

We also look at AGN fractions using a WHAN diagram (Fig. \ref{WHAN}), described by \cite{CidFernandes2010,CidFernandes2011} which unlike the BPT can identify `fake' AGN (i.e. retired galaxies) from those with weakly active nuclei from the BPT LINER region. Using the SDSS control sample we find that 46.0\% of the general galaxy population is made up of star forming galaxies, 27.2\% passive galaxies, 19.3\% strong AGN and 7.4\% weak AGN.

We find that the H$\delta$ strong sample is comprised of 59.7\%$\pm$6.1 star forming galaxies, 15.1\%$\pm$3.1 passive galaxies, 16.4\%$\pm$3.2 sAGN and 8.8\%$\pm$2.3 wAGN. The E+A sample is found to be comprised of 54.2\%$\pm$10.6 star forming galaxies, 27.1\%$\pm$7.5 passive galaxies, 12.5\%$\pm$5.1 sAGN and 8.3\%$\pm$4.2 wAGN. These fractions are similar to those found in the SDSS control sample, however, we witness a decline in passive galaxies and slight enhancement of star formers within the H$\delta$ strong sample compared to SDSS galaxies, which is to be expected.

\section{Discussion}
\subsection{Links to Previous Studies}

We have discussed in section 1 how different studies have used different selection methods to select post-starburst galaxies. There are two main methods for selection; the most common selecting on the lack of [OII] emission and the presence Balmer absorption whilst the other selects on H$\alpha$ emission, as well as [OII] emission and Balmer absorption.

\cite{Poggianti1999,Poggianti2009} find that there are two populations of E+As when they select using [OII] emission and H$\delta$ Balmer absorption; these two populations can be classed as red and blue. We have seen from Fig. \ref{colour_mag} and Fig. \ref{colour_hist} that the H$\delta$ galaxies peak in the blue cloud, the E+As peak in the green valley and the pure E+As peak in the red sequence. This result matches with that of \cite{Hogg2006} who place H$\alpha$ selected E+As within the red sequence. This means the cut in H$\alpha$ is removing galaxies with residual star formation, therefore, we are intrinsically selecting the red E+As mentioned in \cite{Poggianti1999} and \cite{Poggianti2009}.

Along with colour we have also tested the morphologies of our three samples. The literature tells us E+As are bulge dominated with an underlying disc component (\citealt{Tran2004}; \citealt{Quintero2004}), similar to the S0 morphology. This was not seen to change with different selection methods mentioned above. We find in Table \ref{morphologies} and Fig. \ref{fracdev} that there are significant fractions in both the elliptical and spiral morphologies for H$\delta$ strong galaxies and E+As. However, we see that pure E+As are predominantly elliptical.

Regardless of what selection method is used in the literature there is a great deal of discrepancy in the environmental preferences of post-starburst galaxies. Table \ref{yang} shows that $\sim$70\% and $\sim$60\% of H$\delta$ strong galaxies and E+As are within the field, this result agrees with the findings of \cite{Zabludoff1996} at low redshifts $z < 0.1$. \cite{Lemaux2016} states that pure E+As have a higher fraction in clusters than E+As selected only on [OII] emission and Balmer absorption, we find the same trend when examining the fractions of post-starburst galaxies in various environments. \cite{Mahajan2013h} find that E+As prefer a weak-group environment, whilst we do see a systematically higher fraction in the weak-group environment compared with rich environments this result is not statistically significant.

\subsection{Duty Cycle of Post-Starburst Galaxies}

In section 3 we find that the colours of H$\delta$ strong galaxies peak in the blue cloud, the E+As in the green valley and the pure E+As in the red sequence (Fig. \ref{colour_mag} and Fig. \ref{colour_hist}). This result is backed up morphologically in section 4. We find that the majority of H$\delta$ strong galaxies have a spiral morphology whereas the E+As have a mix of spiral and elliptical morphologies . The majority of pure E+As display an elliptical morphology (Table \ref{morphologies}). This is confirmed again by the light profiles of our samples (Fig. \ref{fracdev}), we find that H$\delta$ strong and E+A galaxies have a disky light profile whereas the pure E+As have a de Vaucouleur light profile, typical of an early-type morphology. This difference in morphology is seen in the radii of our galaxies (Fig. \ref{radii} and Fig.\ref{shen_mag}), in which the H$\delta$ strong and E+A galaxies follow the late-type relationship found by \cite{Shen2003} and the pure E+As follow the early-type relationship. The lack of merging morphologies suggests that major mergers do not play an important role in the post-starburst phase.

We speculate that the H$\delta$ strong galaxies, E+A and pure E+As are part of an evolutionary sequence in which star forming galaxies are quenched into blue disky H$\delta$ strong galaxies. They are then quenched further into green valley E+As before, finally, quenching fully into red elliptical pure E+As. If this is true, how does the morphology transition from disk to elliptical? This could be due to a fading disk with time as young stars die or possibly due to any mergers. We add here that we witness no evidence for mergers in the morphologies (Table \ref{morphologies}) so the trigger would have to be a minor merger or a mergerless interaction.

Another factor we consider is environment, although we have low-number statistics, we see a higher proportion of pure E+As in denser environments than the H$\delta$ strong and E+A galaxies. Whilst in the H$\delta$ strong and E+A galaxies we see a strong preference for a field environment. We note this is not a migration of galaxies from the field to a cluster environment as the migration time scale is much longer than that of the evolution of our galaxies. Instead, this may point to an accelerated evolution from H$\delta$ strong galaxies to pure E+As in dense environments (\citealt{Hatch2011d}). A dense environment would enhance processes such as interactions and minor mergers. Also the hot atmosphere of the intracluster medium would help quench and fade the disk by starving the galaxy of the fuel to replenish its cold gas supply.

We see from looking at AGN fractions using BPT and WHAN diagnostics that there is a higher fraction of AGN in the H$\delta$ strong sample compared with the E+As. This could suggest that E+As are the product of H$\delta$ strong galaxies that have been quenched by AGN feedback, however, as the E+A cut selects against line emission this is purely speculative and we have no solid evidence that AGN are important in quenching star formation after a burst.

\section{Conclusion}
The post-starburst phase in galaxy evolution is an important and significant link between star-forming discs and quiescent early-type galaxies. Throughout the literature there are several different selection techniques. We split these different techniques into two categories; those that select on the presence of Balmer absorption and the absence of [OII] emission and those which also select against H$\alpha$ emission. There are advantages and disadvantages in using either method, one includes those galaxies that have H$\alpha$ due to the presence of AGN and the other excludes dusty starburst galaxies. In this work we have explored how the different selection techniques can affect the results found in colour, morphology, environment and AGN fraction. Here we list our principal findings and suggest what we feel is an ideal selection method:

\begin{enumerate}
	\item By locating our samples on a colour magnitude diagram and colour histogram, we have been able to find where each population peaks. We find that H$\delta$ strong galaxies peak in the blue cloud, E+A galaxies in the green valley and pure E+As in the red sequence.
	
	\item Using the GZ catalogue we find that H$\delta$ strong galaxies are predominantly spiral in morphology with a fraction of 54\%. We find that E+As have approximately an even split between the elliptical and spiral morphologies and pure E+As have an elliptical fraction of 81\%. This results is confirmed when we look at the $fracdev$ parameter which shows the H$\delta$ strong galaxies and E+As have a disk morphology and the pure E+As have an early-type morphology. This is again demonstrated when we look at the radii of our galaxies, we see that the H$\delta$ strong galaxies and E+As follow the late-type trend found in \cite{Shen2003}, whilst the pure E+As follow the early-type trend.
	
	\item We find that the majority of H$\delta$ strong galaxies and E+A galaxies have a strong preference for the low density environment, however, there is an even split between the low and high density environments within the pure E+A population. We see that there is a higher fraction of pure E+As in denser environments than the other two populations.
	
	\item In this study we have also tested the AGN fractions of our samples by using BPT and WHAN diagrams. Both diagrams suggest that the emission lines of the H$\delta$ strong and E+A populations are excited by star formation and not AGN activity. The AGN fraction of the H$\delta$ strong galaxies is typical of the general galaxy population, so provides no evidence that AGN play a role in quenching star-formation in post-starburst galaxies. However, we see a significant decrease in AGN fraction compared with the general population which could be a consequence of AGN fading into the E+A phase assuming AGN feedback leads to the E+A phase. However, we note that the E+A sample is selected to not have strong emission lines (albeit in [OII]) and so it is not possible to draw strong conclusions.
	
\end{enumerate}

We postulate that different selection criteria allow us to probe different stages of post-starburst galaxies. After the initial starburst has ended the galaxies are seen as blue spirals/disks with a young stellar spectra, indicated by deep Balmer absorption features. However, $\sim$65\% of these disks are still forming stars as seen by their [OII] and/or H$\alpha$ emission lines. If we cut on [OII] emission (and therefore star formation) we see a population of green valley disk galaxies suggestive of the next stage in the post-starburst evolution as galaxies quench towards the red sequence. If we make a further cut on star formation using a lack of H$\alpha$ emission then we find a population of early-type red sequence galaxies. These are galaxies that still show deep Balmer absorption but with no ongoing star formation and represent the final post starburst phase before the underlying stellar population becomes dominated by old stars.

The lack of any significant enhancement in the merger or AGN fractions for the post-starbursts indicates that, while we cannot rule out a scenario whereby the initial starburst was triggered by a major merger and then initially quenched by an AGN, it is not these processes that continue to quench the galaxies down to full quiescence. One puzzle in our proposed evolutionary sequence is how the morphologies have changed from disky to early-type without evidence of major merging. We speculate that this morphological evolution could be due to the secular fading of the star-forming disk leading to a bulge dominated system or to a series of minor interactions not detected in the imaging. There also seems to be a suggestion that the passive, pure E+As are more prevalent in denser environments, which perhaps suggests that the quenching that follows a starburst proceeds more quickly in dense environments perhaps due to the enhancement of minor interactions in groups or the removal of a cold gas supply by the hot intracluster medium of the most massive systems.

\section*{Acknowledgements}

JPS gratefully acknowledges support through the Hintze Research Fellowship. 

Funding for the SDSS and SDSS-II has been provided by the Alfred P. Sloan Foundation, the Participating Institutions, the National Science Foundation, the U.S. Department of Energy, the National Aeronautics and Space Administration, the Japanese Monbukagakusho, the Max Planck Society, and the Higher Education Funding Council for England. The SDSS Web Site is http://www.sdss.org/.

The SDSS is managed by the Astrophysical Research Consortium for the Participating Institutions. The Participating Institutions are the American Museum of Natural History, Astrophysical Institute Potsdam, University of Basel, University of Cambridge, Case Western Reserve University, University of Chicago, Drexel University, Fermilab, the Institute for Advanced Study, the Japan Participation Group, Johns Hopkins University, the Joint Institute for Nuclear Astrophysics, the Kavli Institute for Particle Astrophysics and Cosmology, the Korean Scientist Group, the Chinese Academy of Sciences (LAMOST), Los Alamos National Laboratory, the Max-Planck-Institute for Astronomy (MPIA), the Max-Planck-Institute for Astrophysics (MPA), New Mexico State University, Ohio State University, University of Pittsburgh, University of Portsmouth, Princeton University, the United States Naval Observatory, and the University of Washington

%%%%%%%%%%%%%%%%%%%%%%%%%%%%%%%%%%%%%%%%%%%%%%%%%%



\bibliographystyle{mnras}
\bibliography{earef} % if your bibtex file is called example.bib

%\begin{thebibliography}{99}
%\end{thebibliography}

%%%%%%%%%%%%%%%%%%%%%%%%%%%%%%%%%%%%%%%%%%%%%%%%%%

%%%%%%%%%%%%%%%%% APPENDICES %%%%%%%%%%%%%%%%%%%%%


%%%%%%%%%%%%%%%%%%%%%%%%%%%%%%%%%%%%%%%%%%%%%%%%%%

% Don't change these lines
\bsp	% typesetting comment
\label{lastpage}
\end{document}